\pdfoutput=1
\documentclass[11pt,twoside,a4paper,cmspaper,final,collab]{cms-tdr}
\def\svnVersion{df19961}\def\svnDate{2025/09/07}\def\cmsCernNoTag{CERN-EP-2025-113}\def\cmsCernDate{\today}\def\cmsMessage{Published in Physics Letters B as \href{http://dx.doi.org/10.1016/j.physletb.2025.139857}{\doi{10.1016/j.physletb.2025.139857}.}}
\begin{document}\cmsNoteHeader{TOP-24-012}

\newcommand{\vectheta}{\ensuremath{\boldsymbol{\theta}}\xspace}
\newcommand{\CP}{\ensuremath{CP}\xspace}
\newcommand{\ctWI}{\ensuremath{c_{\PQt\PW}^{\mathrm{I}}}\xspace}
\newcommand{\ctZI}{\ensuremath{c_{\PQt\PZ}^{\mathrm{I}}}\xspace}
\newcommand{\OtWI}{\ensuremath{\mathcal{O}_{\PQt\PW}^{\mathrm{I}}}\xspace}
\newcommand{\OtGI}{\ensuremath{\mathcal{O}_{\PQt\HepParticle{G}{}{}}^{\mathrm{I}}}\xspace}
\newcommand{\OtZI}{\ensuremath{\mathcal{O}_{\PQt\PZ}^{\mathrm{I}}}\xspace}
\newcommand{\tZq}{\ensuremath{\PQt\PZ\PQq}\xspace}
\newcommand{\ttZ}{\ensuremath{\ttbar\PZ}\xspace}
\newcommand{\ttW}{\ensuremath{\ttbar\PW}\xspace}
\newcommand{\ttH}{\ensuremath{\ttbar\PH}\xspace}
\newcommand{\ttA}{\ensuremath{\ttbar\PGg}\xspace}
\newcommand{\WZ}{\ensuremath{\PW\PZ}\xspace}
\newcommand{\ZZ}{\ensuremath{\PZ\PZ}\xspace}
\newcommand{\pp}{\ensuremath{\Pp\Pp}\xspace}
\newcommand{\mZ}{\ensuremath{m_{\PZ}}\xspace}
\newcommand{\lZp}{\ensuremath{\Pell^{\PZ+}}\xspace}
\newcommand{\lZm}{\ensuremath{\Pell^{\PZ-}}\xspace}
\newcommand{\lW}{\ensuremath{\Pell^{\PW}}\xspace}
\newcommand{\abseta}{\ensuremath{\abs{\eta}}\xspace}
\newcommand{\wSM}{\ensuremath{w_{\text{SM}}}\xspace}

\newcommand{\ctWIdown}{-2.7}
\newcommand{\ctWIup}{2.5}
\newcommand{\ctZIdown}{-0.2}
\newcommand{\ctZIup}{2.0}

\newcommand{\ctWIdownlinear}{-4.5}
\newcommand{\ctWIuplinear}{1.2}
\newcommand{\ctZIdownlinear}{1.0}
\newcommand{\ctZIuplinear}{8.3}

\ifthenelse{\boolean{cms@external}}{%
    \providecommand{\cmsleft}{upper\xspace}
    \providecommand{\cmsLeft}{Upper\xspace}
    \providecommand{\cmsright}{lower\xspace}
}{%
    \providecommand{\cmsleft}{left\xspace}
    \providecommand{\cmsLeft}{Left\xspace}
    \providecommand{\cmsright}{right\xspace}
}

\cmsNoteHeader{TOP-24-012}
\title{Search for \texorpdfstring{\CP}{CP} violation in events with top quarks and \texorpdfstring{\PZ}{Z} bosons at \texorpdfstring{$\sqrt{s}=13$}{sqrt(s)=13} and \texorpdfstring{13.6\TeV}{13.6 TeV}}

\abstract{A search for the violation of the charge-parity (\CP) symmetry in the production of top quarks in association with \PZ bosons is presented, using events with at least three charged leptons and additional jets. The search is performed in a sample of proton-proton collision data collected by the CMS experiment at the CERN LHC in 2016--2018 at a center-of-mass energy of 13\TeV and in 2022 at 13.6\TeV, corresponding to a total integrated luminosity of 173\fbinv. For the first time in this final state, observables that are odd under the \CP transformation are employed. Also for the first time, physics-informed machine-learning techniques are used to construct these observables. While for standard model (SM) processes the distributions of these observables are predicted to be symmetric around zero, \CP-violating modifications of the SM would introduce asymmetries. Two \CP-odd operators \OtWI and \OtZI in the SM effective field theory are considered that may modify the interactions between top quarks and electroweak bosons. The obtained results are consistent with the SM prediction within two standard deviations, and exclusion limits on the associated Wilson coefficients of $\ctWIdown < \ctWI < \ctWIup$ and $\ctZIdown < \ctZI < \ctZIup$  are set at 95\% confidence level. The largest discrepancy is observed in \ctZI where data is consistent with positive values, with an observed local significance with respect to the SM hypothesis of 2.5 standard deviations, when only linear terms are considered.}

\hypersetup{%
pdfauthor={CMS Collaboration},%
pdftitle={Search for CP violation in events with top quarks and Z bosons at sqrt(s)=13 and 13.6 TeV},%
pdfsubject={CMS},%
pdfkeywords={CMS, top quark, CP violation, EFT, Z boson}}

\maketitle

\section{Introduction}

The violation of the charge-parity (\CP) symmetry is one of the necessary conditions to explain
the baryon asymmetry (BAU) observed in the universe~\cite{Sakharov:1967dj}. In the standard model (SM), \CP violation occurs thanks to the
imaginary phases in quark and neutrino flavor mixing matrices of the weak interaction. The magnitude associated to these SM sources is however insufficient to account for the BAU, motivating
the search for additional sources of \CP violation introduced by interactions beyond the SM (BSM).
In cases where the BSM scale is above the accessible energy in an experiment, its effect can still influence the results.
Phenomenologically, these effects introduced can be parameterized by an effective field theory (EFT).
The SM EFT~\cite{Degrande:2012wf} introduces a number of operators that are odd under \CP transformations and
that could give rise to \CP violating phenomena.

\begin{figure}[!ht]
\centering
\includegraphics[width=0.47\textwidth]{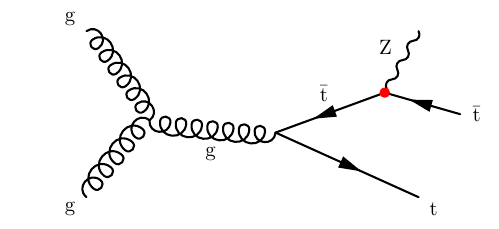}%
\hfill%
\includegraphics[width=0.47\textwidth]{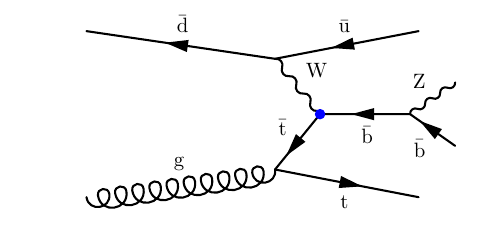}
\caption{%
    Example Feynman diagrams for \ttZ (\cmsleft) and \tZq (\cmsright) production with vertices that can be
    modified by \ctZI (\ctWI) highlighted in red (blue) circle.
}
\label{fig:feyn_diagrams}
\end{figure}

In this letter, we explore the presence of \CP violation in the associated production of a \PZ boson with a top quark pair (\ttZ) or with a single top quark (\tZq),
with example Feynman diagrams shown in Fig.~\ref{fig:feyn_diagrams}. The highlighted couplings could be modified by the \OtWI and \OtZI dimension-6 operators,
as defined in Ref.~\cite{AguilarSaavedra:2018nen}, that correspond to the imaginary part of the electroweak dipole operators.
The strength of the operators' contribution is modulated by its associated Wilson coefficients, \ctWI and \ctZI.
The inclusive
and differential production cross sections for these two processes have been
measured by the ATLAS and CMS Collaborations~\cite{ATLAS:2023eld, ATLAS:2020bhu, CMS:2019too, CMS:2021ugv, CMS:2024mke}, reporting results consistent with the SM expectations in
observables that are invariant under \CP transformations. References~\cite{CMS:2024mke, CMS:2019too} report mild excesses in the \ttZ production rate, within two standard deviations of the SM prediction. The \OtGI operator may also introduce \CP-violating effects in these processes, however it was checked that the sensitivity to this operator is smaller than the one obtained in top quark pair production~\cite{CMS:2022voq}.

We exploit observables whose sign flips when evaluated on the \CP transform of any given event, \ie, they are odd under \CP transformations.
To construct such observables, we employ physics-informed machine-learning techniques~\cite{DBLP:journals/corr/abs-1711-10561}, which allow us
to train algorithms to solve supervised learning tasks, while respecting any given laws of physics.
In particular, we consider neural networks that are equivariant under the \CP symmetry~\cite{Cruz:2024grk}.
The SM prediction follows an even distribution under our \CP-odd observables, whereas \CP-violating effects in the signal processes would
result in an asymmetry. Although the decay channels considered do not allow for a complete reconstruction of all final-state objects,
the observables are marginalized over the unresolved degrees of freedom, maintaining the aforementioned symmetry properties.
This analysis presents the first search in top quark production in association with gauge bosons that exploits \CP-odd observables and is therefore
sensitive to new physics effects that do not modify the total cross section of the studied processes. It is also the
first search for \CP violation where physics-informed machine learning is applied. Physics-informed machine learning allows us to incorporate our knowledge on the expected signature from the \CP-violating signal, which enhances the robustness and sensitivity of the search. The training of these algorithms converges to the optimal observables, exploiting the modification of the signal kinematic features in the signal introduced by the BSM interaction as well as features discriminating signal from the main backgrounds.

The distribution of \CP-odd observables in \ttZ and \tZq production is measured with proton-proton (\pp)
collisions at $\sqrt{s}=13$ (13.6)\TeV collected in 2016--2018 (2022) with the CMS experiment at the CERN LHC.
The 2016--2018 (2022) data correspond to an integrated luminosity of 138 (34.7)\fbinv.
We analyze a sample of events with at least three charged leptons (electrons and/or muons) with additional jets, targeting leptonic decays of the \PZ boson
and at least one of the \PW bosons from the top quark decays. We obtain confidence intervals for
\ctWI and \ctZI using the observed yields in the distribution of the \CP-odd observables, reaching
comparable sensitivity to other searches exploring these operators using \CP-even
observables~\cite{ATLAS:2023eld, ATLAS:2024hmk, CMS:2021klw, CMS:2022lmh} and to a search for \CP violation in $t$-channel single top quark production performed by ATLAS~\cite{ATLAS:2022vym}.
This search sets for the first time limits on the interference between the \ctZI and SM contributions.
Tabulated results are provided in the HEPData record for this analysis~\cite{hepdata}.

\section{The CMS detector and object reconstruction}
\label{sec:detector}

The CMS apparatus~\cite{CMS:2008xjf,CMS:2023gfb} is a multipurpose, nearly hermetic detector, designed to
trigger on~\cite{CMS:2020cmk, CMS:2016ngn, CMS:2024psu} and identify electrons, muons, photons, and (charged
and neutral) hadrons~\cite{CMS:2020uim, CMS:2018rym, CMS:2014pgm}.
A ``particle-flow'' (PF) algorithm~\cite{CMS:2017yfk} aims to reconstruct all individual particles in
an event, combining information provided by the all-silicon inner tracker and by the crystal electromagnetic (ECAL)
and brass-scintillator hadron calorimeters, operating inside a 3.8\unit{T} superconducting solenoid, with data
from the gas-ionization muon detectors embedded in the flux-return yoke outside the solenoid. The reconstructed
particles are used to build \PGt leptons, jets, and missing transverse momentum~\cite{CMS:2018jrd,CMS:2016lmd,CMS:2019ctu}.

Electrons (muons) are reconstructed with absolute pseudorapidity  $\abseta<2.5$ (2.4), corresponding to the geometrical
acceptance of the silicon strip tracker (muon system). 
The algorithms described in Refs.~\cite{CMS:2020uim,CMS:2020mpn,CMS:2023dsu} are used to efficiently select electrons and muons
with transverse momentum $\pt>15\GeV$  produced in the
prompt decays of \PW and \PZ bosons, or prompt \PGt leptons, while
rejecting those that are coming from other sources. 

Jets are clustered using the anti-\kt algorithm~\cite{Cacciari:2008gp,Cacciari:2011ma} with a distance parameter of 0.4 and their  momentum is determined as the vectorial sum of all particle momenta
in the jet, and is found from simulation to be, on average, within 5 to 10\% of the true
particle-level jet momentum over the entire \pt spectrum and detector acceptance. Additional \pp interactions within the same or nearby bunch crossings (pileup) can contribute
additional tracks and calorimetric energy depositions, increasing the apparent jet momentum. To mitigate this effect, tracks identified to be originating from pileup
vertices are discarded. For events recorded at $\sqrt{s}=13\TeV$, an offset correction is applied to correct for remaining contributions~\cite{CMS:2020ebo}, while for events at
$\sqrt{s}=13.6\TeV$ the pileup-per-particle identification algorithm~\cite{CMS:2020ebo, Bertolini:2014bba} is used.
Jet energy corrections are derived from simulation studies so that the average measured energy of jets becomes identical to that of particle level jets.
In situ measurements of the momentum balance in dijet, photon+jet, {\PZ}+jet, and multijet events are used to determine any
residual differences between the jet energy scale (JES) in data and in simulation, and appropriate corrections are made~\cite{CMS:2016lmd}.
The missing transverse momentum vector \ptvecmiss is computed as the negative vector sum of the transverse momenta of all the PF candidates in an event,
and its magnitude is denoted as \ptmiss~\cite{CMS:2019ctu}.
The \ptvecmiss is modified to account for corrections to the energy scale of the reconstructed jets in the event.

Jets in the event are required to have $\pt>25\GeV$ and $\abseta<5 (2.4)$ for events at
$\sqrt{s}=13$ (13.6)\TeV. The reason for the more restrictive \abseta selection in $\sqrt{s}=13.6\TeV$ data
stems from the mismodeling of detector noise during this run period, which is more challenging for PF candidates with $\abseta>2.5$ as their reconstruction relies exclusively on calorimeter information. Jets with $\abseta<2.4$ originating from the hadronization of bottom quarks (\PQb jets) are identified
by the \textsc{DeepJet} algorithm~\cite{Bols_2020, CMS-DP-2023-012, CMS-DP-2023-005}. Two working points are considered, corresponding to selection
efficiencies of 84\% (``loose'') and 70\% (``tight'') for \PQb jets, 11\% and 1\% for light-quark or gluon jets,
and 50\% and 15\% for charm quark jets, respectively.

Events of interest are selected using a two-tiered trigger system. The first level, composed of custom hardware processors, uses information from
the calorimeters and muon detectors to select events at a rate of around 100\unit{kHz} within a fixed latency of about 4\mus~\cite{CMS:2020cmk}. The second
level, known as the high-level trigger, consists of a farm of processors running a version of the full event reconstruction software optimized
for fast processing, and reduces the event rate to a few kHz before data storage~\cite{CMS:2016ngn, CMS:2024psu}.

\section{Data and simulated samples}
\label{sec:datasets}

This analysis uses data collected
with a combination of single-, double-, and triple-lepton triggers, which have an efficiency larger than 98\% for events containing at least
three leptons passing the requirements described in Section~\ref{sec:event_selection}.

We employ different Monte Carlo event generators to produce samples of simulated events, which are used to determine the SM expectation
of the different observables used in the analysis. The \MGvATNLO v2.6.5 or v2.9.18~\cite{Alwall:2014hca, Alwall:2007fs, Artoisenet:2012st, Frederix:2012ps}
generator is used for the production of simulated samples of \tZq, \ttZ, the production of a top quark-antiquark pair in association with a \PW or a Higgs boson, or with a photon (\ttW, \ttH, \ttA),
of a single top quark with a \PW and a \PZ boson, and the production of four top quarks at next-to-leading order (NLO) accuracy.
The \ttW, \ttH, and \ttA samples are generated with one additional parton in the matrix element calculations.
We simulate events involving the production of a pair of massive gauge bosons, \WZ and \ZZ, using \POWHEG v2~\cite{Nason:2004rx,Frixione:2007vw,Alioli:2010xd} at NLO accuracy.
Other minor backgrounds, such as the triple boson
production or the production of top quarks in association with two gauge bosons are simulated at leading order (LO) using the same version of \MGvATNLO described above.
All samples are normalized to the most accurate theoretical calculations available, namely approximate next-to-next-to-NLO for \ttZ from Ref.~\cite{Kidonakis:2024lht} and
NLO for \tZq, obtained using \MGvATNLO.

The contribution from the \OtWI and \OtZI operators is estimated using simulated samples of the \ttZ and \tZq processes, produced with the \MGvATNLO v2.9.18 generator at LO accuracy,
using the \textsc{dim6top} model~\cite{AguilarSaavedra:2018nen} to incorporate the EFT effects.
We consider scenarios where only \ctWI and \ctZI take values different from zero.
The decay of the top quarks into $\PW\PQb$ and the subsequent decay of the \PW bosons, as well as that of the \PZ bosons, is simulated using \MGvATNLO to properly account for
spin correlations of the final-state particles. The EFT effects in the production are incorporated through per-event weights, using the approach described in Refs.~\cite{Mattelaer:2016gcx,CMS:2023xyc}.
The EFT effects in the decay of the top quarks into $\PW\PQb$ were found to be negligible in this analysis.
In this way, each simulated event receives a parametric weight $w$
given by
\begin{equation}\label{eq:eft_parameterization}
    w = \wSM + \sum_{i}  c^{\mathrm{I}}_i l_i + \sum_{ij} c^{\mathrm{I}}_i c^{\mathrm{I}}_j q_{ij},
\end{equation}
where \wSM is the SM contribution and the sums run over $\PQt\PW$ and $\PQt\PZ$. The terms
involving $l_i$ ($q_{ij}$) represent the linear (quadratic) contributions of the \ctWI and \ctZI operators to the event weight.
They represent the
contribution from the interference between SM and BSM diagrams (from purely BSM contributions).
We refer to the former contributions as the ``linear'' contributions and the latter as the ``quadratic''
contributions.

All simulations use the NNPDF3.1~\cite{ref:nnpdf} set of parton distribution functions (PDFs). The generators are then interfaced with \PYTHIA v8.240~\cite{Sjostrand:2014zea} (2016--2018 data) or v8.306~\cite{Bierlich:2022pfr} (2022 data)
for showering, decay, hadronization, and underlying event simulation,
using the CP5 tune~\cite{ref:cp5}. To avoid double-counting between the matrix element generation in \MGvATNLO and the parton shower, the FxFx~\cite{Frederix:2012ps}
(MLM~\cite{Alwall:2007fs}) matching scheme is used for processes simulated at NLO (LO) accuracy with additional partons in the matrix element calculations. The CMS detector response is simulated using \GEANTfour~\cite{Agostinelli:2002hh}.
The effect of pileup interactions is taken into account by adding simulated minimum-bias interactions to the simulated hard-scattering event.

\section{Event selection and background estimation}
\label{sec:event_selection}

We select events with at least three reconstructed leptons passing the selection mentioned in Section~\ref{sec:detector}
and with \pt larger than 25, 15, and 15\GeV, respectively, in order for the leptons to be well above the threshold of the trigger selection. Events are also required to have a pair
of opposite-sign same-flavor leptons whose invariant mass is less than 15\GeV away from the \PZ boson mass $\mZ=91.2\GeV$~\cite{ParticleDataGroup:2024cfk}.
In addition, only events with at least two reconstructed jets of which at least one
passes the tight \PQb tagging requirement are selected.

Events passing this selection are dominated by the signal processes, \ttZ and \tZq production,
but have a subdominant contribution from  \WZ and \ZZ production, which we refer to collectively
as ``diboson''. Events with nonprompt leptons passing our selection and with photons undergoing
conversions to electrons in their interaction with the detector material~\cite{CMS:2015xaf} also have a sizable
contribution to the events passing this selection. Residual contributions
from other processes listed in Section~\ref{sec:datasets} remain, which we refer collectively as ``rares''.

The contribution from each process, with the exception of events containing nonprompt leptons, is
estimated using samples of simulated events. Simulated events are corrected to account for
small discrepancies observed between data and simulations in the lepton and jet selection efficiency,
\PQb tagging efficiency, JES and jet energy resolution (JER), using dedicated calibration
methods~\cite{CMS:2017wtu, CMS-DP-2023-012, CMS-DP-2023-005, CMS:2010svw, CMS:2016lmd},
implemented through scale factors (SFs) used to weight event samples or by varying the relevant quantities.

The contribution from processes with nonprompt leptons is estimated using the misidentification probability
method described in Ref.~\cite{CMS:2020mpn}. The method relies on events satisfying the nominal selection
criteria, except that at least one of the leptons is required to pass looser selection criteria and fail the lepton nominal selection criteria. This sample
of events is enriched in processes with nonprompt leptons and is weighted with a suitable transfer
factor to estimate the contribution from this process to the nominal selection. We determine the transfer factor
from data in a dedicated sample of events with a single lepton, which is enriched in multijet events containing a nonprompt lepton.

The accuracy of the SM background prediction was checked in dedicated validation regions that are defined by inverting the
selection criteria applied to the number of \PQb-tagged jets and/or the requirement of an opposite-sign same-flavor
lepton pair consistent with the \PZ boson decay. The number of observed events in these regions
agrees well with the SM predictions within the associated uncertainties.

\section{Construction of \texorpdfstring{\CP-odd}{CP-odd} observables}
\label{sec:CP_observables}

Operators in the SM EFT give rise to two kinds of phenomena, corresponding to the last two terms in Eq.~\eqref{eq:eft_parameterization}, which are linked
to the Feynman amplitudes contributing to the signal processes.
These terms correspond to the interference between SM and BSM Feynman amplitudes (``linear'')  and to purely BSM amplitudes
(``quadratic''), which are proportional to $1/\Lambda^2$ and $1/\Lambda^4$, respectively. Here $\Lambda$ represents
the new physics energy scale and is set in the following to 1\TeV, as a convention. We note that, in the EFT framework, we are not simultaneously sensitive to the Wilson coefficients and $\Lambda$. Linear terms are the leading contribution
in powers of $1/\Lambda$, whereas quadratic terms correspond to higher-order corrections. When considering only dimension-6 operators,
this construction is only consistent up to the linear term, as the pure BSM
contribution is at the same order as the dimension-8 operators, which are not considered in this analysis.
The linear term gives rise
to \CP-odd contributions, whereas both the SM and the quadratic term give rise
to \CP-even contributions, making searches based on the linear contributions more robust
to theoretical and experimental unknowns.
Since quadratic contributions may still give effects observable with our current experimental sensitivity,
we consider separately the case where only the linear contribution is present and the case where both the linear and quadratic contributions are present.

Several interesting processes are present when studying the effect of \CP-odd operators and final states
that are \CP eigenstates, such as \ttZ production. Linear effects cancel out when considering \CP-even observables. This is due to the fact that the \CP transformation changes the sign of the linear contribution. Since each event and its \CP-transform contribute to the same bin of a \CP-even observable, this results in a net linear contribution equal to zero.
As a consequence, for instance, linear contributions do not affect the total cross section of the studied processes.
Instead, linear effects modify the distribution of \CP-odd observables, introducing a characteristic asymmetry. In addition, \CP-conserving
processes are distributed symmetrically under \CP-odd observables.
Although the premise that we are considering a \CP eigenstate does not always hold at the LHC (for instance, it is not the case for \tZq production,
which is dominated by the $\PQq\Pg$ partonic initial state),
these properties are still present in studies considering other states that are not \CP eigenstates, as it often happens
in similar observables described in the literature~\cite{Panico:2017frx, Bar-Shalom:2024dav, Afik:2022bcz}.
We have verified that these symmetry properties are also present in the observables we construct, as shown in Section~\ref{sec:fits}.

We build \CP-odd observables making use of two \CP-equivariant neural networks~\cite{Cruz:2024grk},
aiming to be sensitive to \ctWI and \ctZI, respectively. Each of these models, $g$, has the property
that, for a given set of input observables $x$, $g(\CP(x))= -g(x)$, where $\CP(x)$ is
the \CP transformation acting on these observables.

To build the input variables to the models,
out of the three leading leptons, we consider the opposite-sign same-flavor pair whose
invariant mass is closest to \mZ and denote the positive (negative) one as
\lZp (\lZm). We refer to the third one as \lW.
As input variables to the network, we consider the 3-momenta of \lZp, \lZm, \lW,
and that of the five leading jets.
For each of the five leading jets, we also
consider a score  that takes values $\lbrace0, 1, 2\rbrace$
for jets that fail the loose,
pass the loose but fail the tight, and pass the tight \PQb tagging working point, respectively.
For events with less than five jets, the variables associated to the missing jets are assigned to zero.
We also consider the two components of \ptvecmiss, the charge sign of \lW,
and an integer variable that takes integer values
0--5 for each of the data-taking periods, allowing the model to account for small differences
in detector conditions and $\sqrt{s}$. The data-taking periods in 2016 and 2022 are each divided into two eras to account for changing detector conditions.

As described in Ref.~\cite{Cruz:2024grk}, one can build our equivariant models, $g_{i}$, with $i\in\lbrace\ctWI,\ctZI\rbrace$ by
the functions $g_{i}(x)=f_{i}(x)-f_{i}(\CP(x))$,
where $f_{i}$ are parametrized using fully connected multilayer perceptrons
with 4 hidden layers with between 160 and 20 neurons per layer and a leaky ReLU
activation function, implemented in the \textsc{pytorch} package~\cite{paszke2019pytorchimperativestylehighperformance}.
Each of the models $g_{i}$ is trained
to minimize the loss functions $L_{i}$, which are similar
to those used in the SALLY method~\cite{Brehmer:2018kdj}, given by
\begin{equation}
    L_{i} = \sum \wSM  \left( \frac{l_{i}}{\wSM} - g_{i}(x)  \right)^2,
\end{equation}
where the definitions of \wSM and $l_{i}$ are discussed in Eq.~\eqref{eq:eft_parameterization}.
The loss function is computed using \ttZ, \tZq, and \WZ simulated events, over which the sum runs.

The likelihood scores as a function of \ctWI and \ctZI, respectively, minimize the aforementioned loss function.
These functions are sufficient summary statistics to probe small values of \ctWI and \ctZI, the regime where the linear contribution dominates.
As a consequence,
the training of $g_{i}$ converges to optimal observables~\cite{Castro:2809469,Diehl:1996wm} for these operators.
We include both signal and \WZ events in the training, which
allows us to obtain observables sensitive to the EFT effects present
in the \ttZ and \tZq signal processes, while also discriminating between
them and \WZ production, which is the leading source of backgrounds.
To be robust against a potential overtraining of the discriminator, we split the sample of simulated events in three parts with 50\%, 25\%, and 25\%
of the simulated events. We use the first part for the training of the networks, the second for the 
selection of hyperparameters and validation of the training. Only the third part, which is not used in the training, is used to estimate the event yields in the signal region.

In order to quantify the improvement of the chosen method with respect to other	approaches, we consider triple products involving the three-momenta of the objects used as inputs to the neural networks, inspired by the observables proposed in Ref.~\cite{Degrande:2021zpv}.  This approach does not give rise to a sizable sensitivity to either of the two operators considered.

\section{Statistical model, systematic uncertainties, and results}
\label{sec:fits}

\begin{figure}[!ht]
\centering
\includegraphics[width=0.47\textwidth]{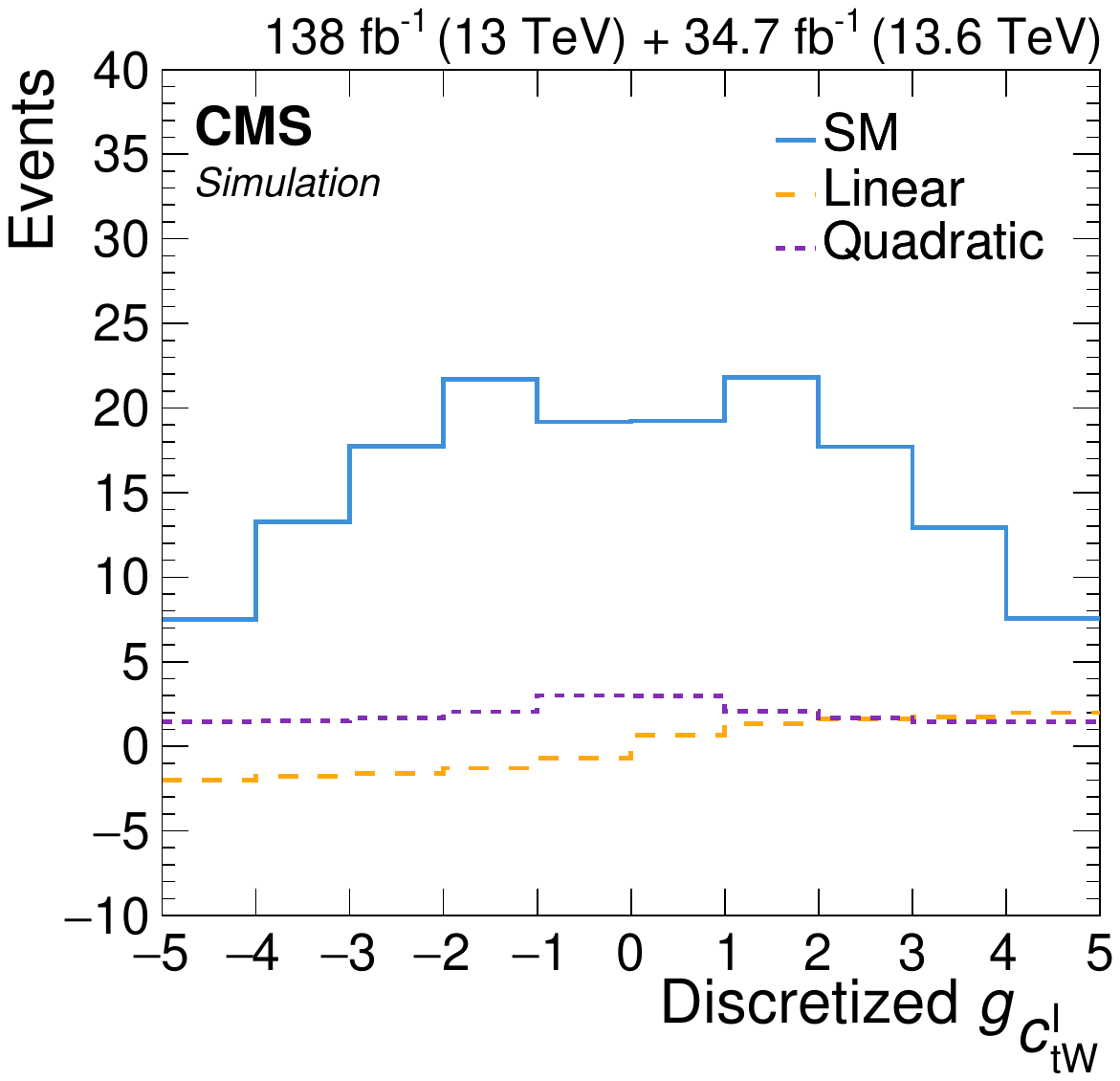}%
\hfill%
\includegraphics[width=0.47\textwidth]{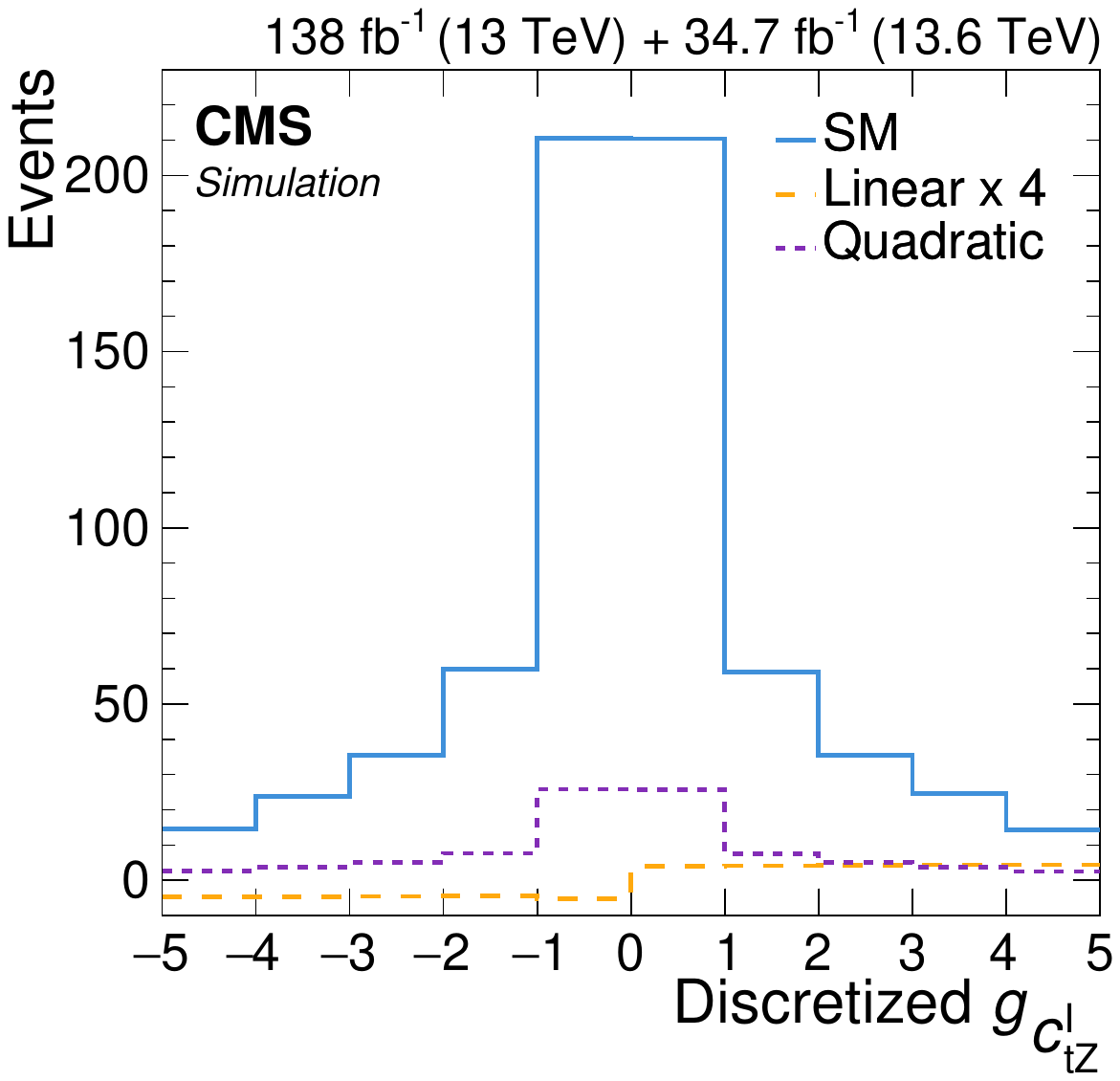}
\caption{%
    \cmsLeft (\cmsright): Distribution of the discretized $g_{\ctWI}$ ($g_{\ctZI}$) for events in the
    \ctWI-like (\ctZI-like) category in \tZq (\ttZ) events. The contributions
    from the SM, linear, and quadratic terms when each Wilson coefficient is set to unity are plotted separately. For better visibility
    the interference contribution in the \ctZI-like category has been scaled by a factor of 4.
}
\label{fig:distribution}
\end{figure}

It has been checked that large values of $g_{\ctWI}$ or $g_{\ctZI}$  correspond to different sets of events, showing that
the discriminators exploit different features in our signals. To exploit that, 
events are classified in two categories according to their $\abs{g_{\ctWI}}$ and $\abs{g_{\ctZI}}$
values as \ctWI-like (\ctZI-like) if $\abs{g_{\ctWI}}>\abs{g_{\ctZI}}$ ($\abs{g_{\ctWI}}<\abs{g_{\ctZI}}$).
In addition, events in the \ctWI-like (\ctZI-like) categories are classified in bins of $g_{\ctWI}$ ($g_{\ctZI}$),
whose edges are symmetrically distributed around zero.
In this analysis, we do not categorize events according to the data-taking era or $\sqrt{s}$.
For the purpose of visualization, we define the discretized $g_{\ctWI}$ ($g_{\ctZI}$) by mapping the bin edges to integer numbers between $-5$ and $5$.
Figure~\ref{fig:distribution} shows the discretized $g_{\ctWI}$ ($g_{\ctZI}$)
for events in the \ctWI-like and \ctZI-like categories in
\tZq and \ttZ events, respectively, illustrating the symmetry properties described in Section~\ref{sec:CP_observables}.

To perform inference on \ctWI and \ctZI, we consider the likelihood function
\begin{equation}
    \mathcal{L}\Big(\ctWI, \ctZI , \vectheta \Big)
    = \prod_{i} \, \mathcal{P}\Big(n_{i} ; \, \ctWI, \ctZI, \vectheta \Big) \, \prod_{k} \, p\Big(\tilde{\theta}_{k} ; \, \theta_{k}\Big),
\label{eq:likelihoodFunction}
\end{equation}
where $n_i$ is the number of observed events in each bin of the distribution and $\mathcal{P}$
is the probability mass function of the associated Poisson distribution. The mean
of the Poisson distribution depends on \ctWI, \ctZI, and \vectheta.
The vector \vectheta comprises a set of nuisance parameters used
to model the effect of systematic
uncertainties, which have a prior given by the function $p$.
We consider the following uncertainty sources by means of weights
or by varying the relevant quantities in the simulated samples used to estimate the predicted yields.
\begin{itemize}
\item The uncertainty in the integrated luminosity is estimated to be 1.6 (1.4)\% for the 2016--2018 (2022) data set~\cite{CMS:LUM-17-003, CMS:LUM-17-004, CMS:LUM-18-002, CMS:LUM-22-001}.
\item Uncertainties in the JES and JER are obtained by varying these quantities in simulated events by one standard deviation. A total of 24 nuisance parameters are used to model the JES uncertainty, introducing different dependencies of the JES as a function of the jet kinematics and the data-taking period. The JER is modeled with 30 nuisance parameters, representing different detector regions and data-taking periods.
\item The uncertainties in the \PQb tagging efficiency and mistagging rate calibrations are assessed by varying the dedicated SFs within their uncertainties. These variations are applied separately for heavy-flavor (\PQb and \PQc quark) jets and light (gluons, \PQu, \PQd, and \PQs quark) jets.
\item The lepton triggering and identification efficiencies are assessed by varying their associated SFs by their uncertainties, separately for electrons and muons. An additional correction and its associated uncertainty are applied to model the so-called \textit{\Lone trigger prefiring}, \ie, a time jitter of the trigger signal in the ECAL~\cite{CMS:2020cmk} and muon~\cite{CMS:2021yvr} systems, which led to a loss of events at the level of 0.5\% or less.
\item The effects due to the uncertainty in the distribution of the number of pileup interactions is estimated by varying the total inelastic \pp cross section used to calculate the number of pileup interactions by 4.6\%.
\item The uncertainty in the nonprompt-lepton contribution is accounted for by variations in the associated transfer factor. This uncertainty includes sources stemming from the measurement of this factor, related to the limited amount of events and the subtraction of the prompt lepton contribution to the measurement. They also include small phase space dependencies of the transfer factor that are not taken into account in its parametrization, mostly dominated by the flavor of the jet producing the nonprompt lepton~\cite{CMS:2020mpn}.
\item The uncertainty in the inclusive \ttZ and \tZq production cross section stems from the accuracy of their theoretical calculations, where we assign a 10\% uncertainty~\cite{LHCHiggsCrossSectionWorkingGroup:2016ypw, CMS:2024mke}. The inclusive diboson production cross section is known and has been measured at a much better precision~\cite{CMS:2021icx,CMS:2020gtj}, but in phase spaces differing from the one in which this search is performed. As a consequence, we take a 20\% normalization uncertainty accounting for this phase space dependence. The normalization uncertainty of the ``rare'' processes and conversions are 50 and 30\%, respectively~\cite{CMS:2020mpn}.
\item Theoretical uncertainties in the kinematic distributions predicted by the simulations stem from three sources: matrix element calculations, PDFs, and parton shower uncertainties. The uncertainties in the matrix element calculations are obtained by shifting independently the renormalization and factorization scales up (down) by a factor of 2 (0.5), respectively. The uncertainty in the PDFs is obtained by using 100 orthogonal eigenvectors of the PDF set, which are used as independent nuisance parameters in the fit. The impact of the parton shower uncertainty is estimated by varying the initial-state and final-state radiation scales up (down) by a factor of 2 (0.5).
\item The number of additional jets generated by the parton shower simulation in diboson samples underestimates the observed distribution in data. This is known from previous measurements~\cite{CMS:2020mpn} and was confirmed in data control samples for both 13 and 13.6\TeV data. We consider an uncertainty whose effects modifies the jet multiplicity distribution by 10--50\% to account for that mismodeling.
\item The uncertainty due to the limited number of simulated events as well as the limited number of data events in the sideband used to estimate the nonprompt-lepton contribution is estimated using the Barlow--Beeston lite method~\cite{Barlow:1993dm}.
\item The LO \tZq samples used to estimate EFT effects predict up to a 30\% smaller acceptance than the NLO ones when weighted to the SM point in the analysis bin where this discrepancy is the largest. The difference between the two samples is used as an additional source of uncertainty in the BSM contribution.
\end{itemize}
In addition, we have evaluated the effect of potential experimental biases
in the symmetry of the $g_{\ctWI}$ and $g_{\ctZI}$ distributions. We considered different scenarios in which the lepton efficiency
and energy scale decrease or increase by up to 10\% as a function of the lepton charge and/or $\eta$. We did not observe
any asymmetry in the SM predictions when smearing the simulations under these scenarios, within the statistical
uncertainty of these predictions.
While these scenarios represent unreasonably biased experimental conditions, this check gives us confidence that potential
experimental biases not taken into account would not introduce a sizable asymmetry.

\begin{figure}[!htp]
\centering
\includegraphics[width=0.47\textwidth]{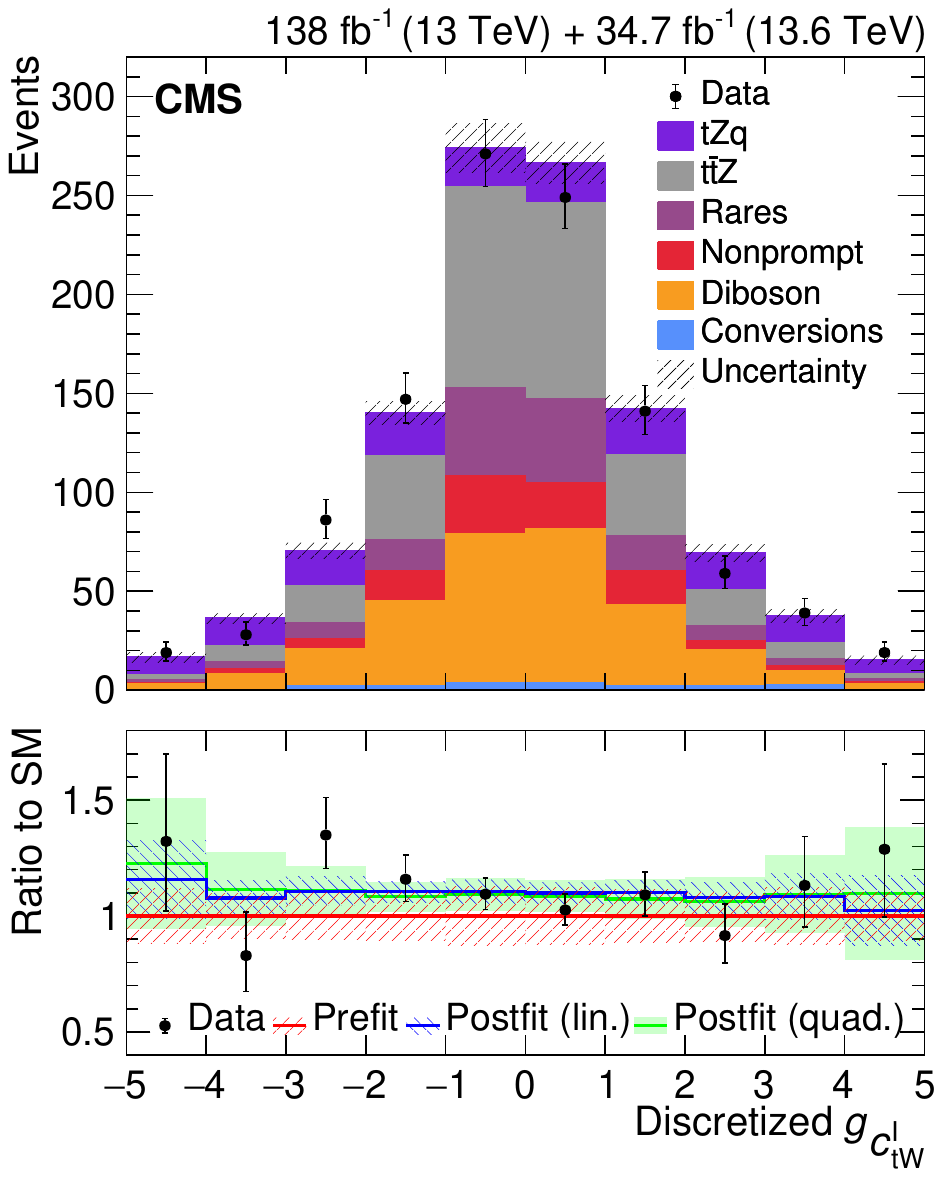}%
\hfill%
\includegraphics[width=0.47\textwidth]{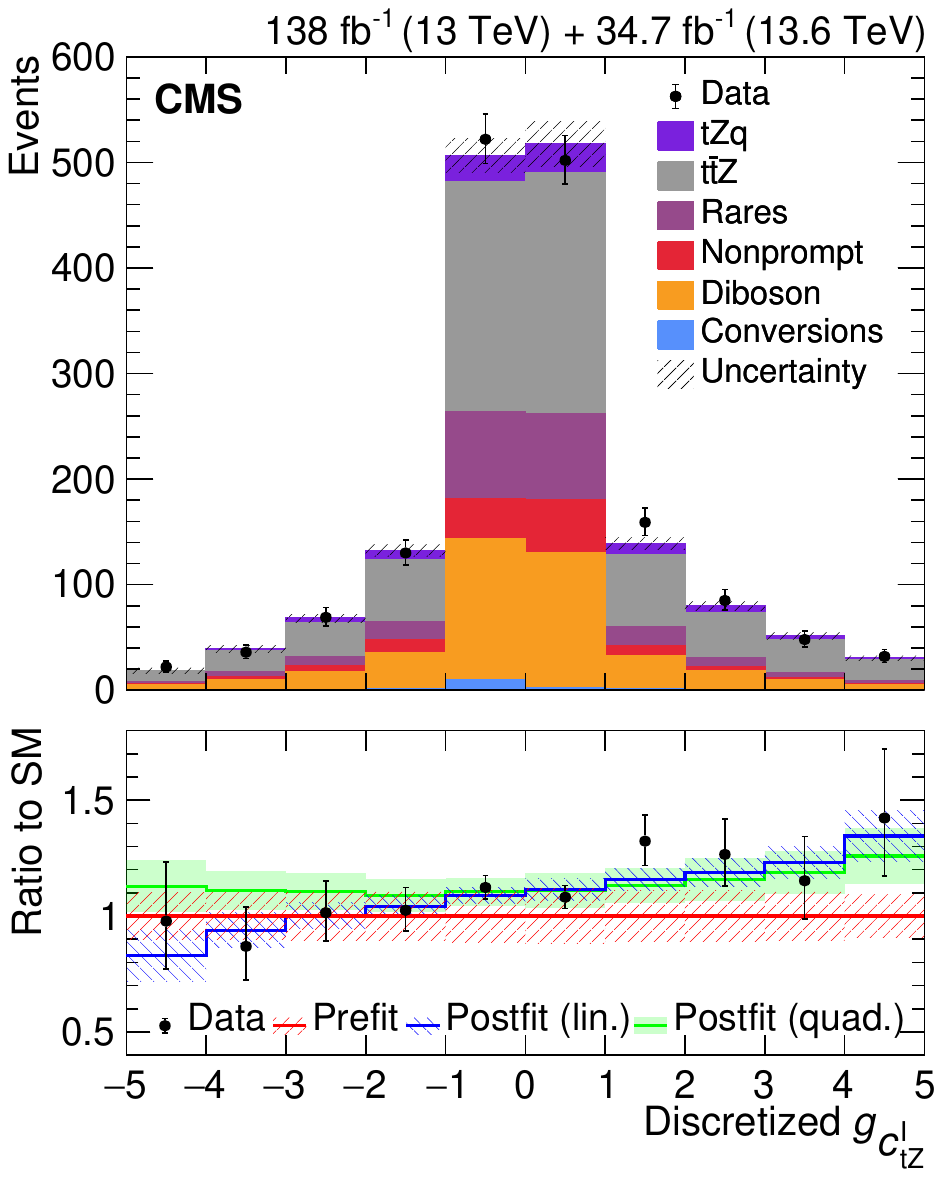}
\caption{%
    \cmsLeft (\cmsright): Distribution of the discretized $g_{\ctWI}$ ($g_{\ctZI}$) for events in the
    \ctWI-like (\ctZI-like) category, compared with the predictions obtained
    when all fit parameters are set to their maximum likelihood value in the linear fit, where \ctWI and \ctZI are  determined simultaneously.
    The lower panels show the ratio between data (black dots), the prediction of the linear (blue line), and quadratic (green line) fits, over the
    prefit value. The red, blue, and green bands show the prefit uncertainty and the postfit uncertainties of the linear and quadratic fits, respectively.
}
\label{fig:stack}
\end{figure}

The distribution of observed events is shown in Fig.~\ref{fig:stack}, classified in the categories
we use as inputs to the likelihood fit. The number of observed events is compared with the
prediction obtained by setting to zero the Wilson coefficient and nuisance parameter values (prefit),
as well as to the values obtained from the maximum likelihood fits described below (postfit).
Small differences between the
prefit and postfit predictions are observed due to a slight asymmetry observed in the $g_{\ctZI}$ distribution,
and to a slight excess on the total number of observed events, consistent
with previous measurements of \ttZ production~\cite{CMS:2024mke, CMS:2019too}.

\begin{figure*}[!htp]
\centering
\includegraphics[width=0.47\textwidth]{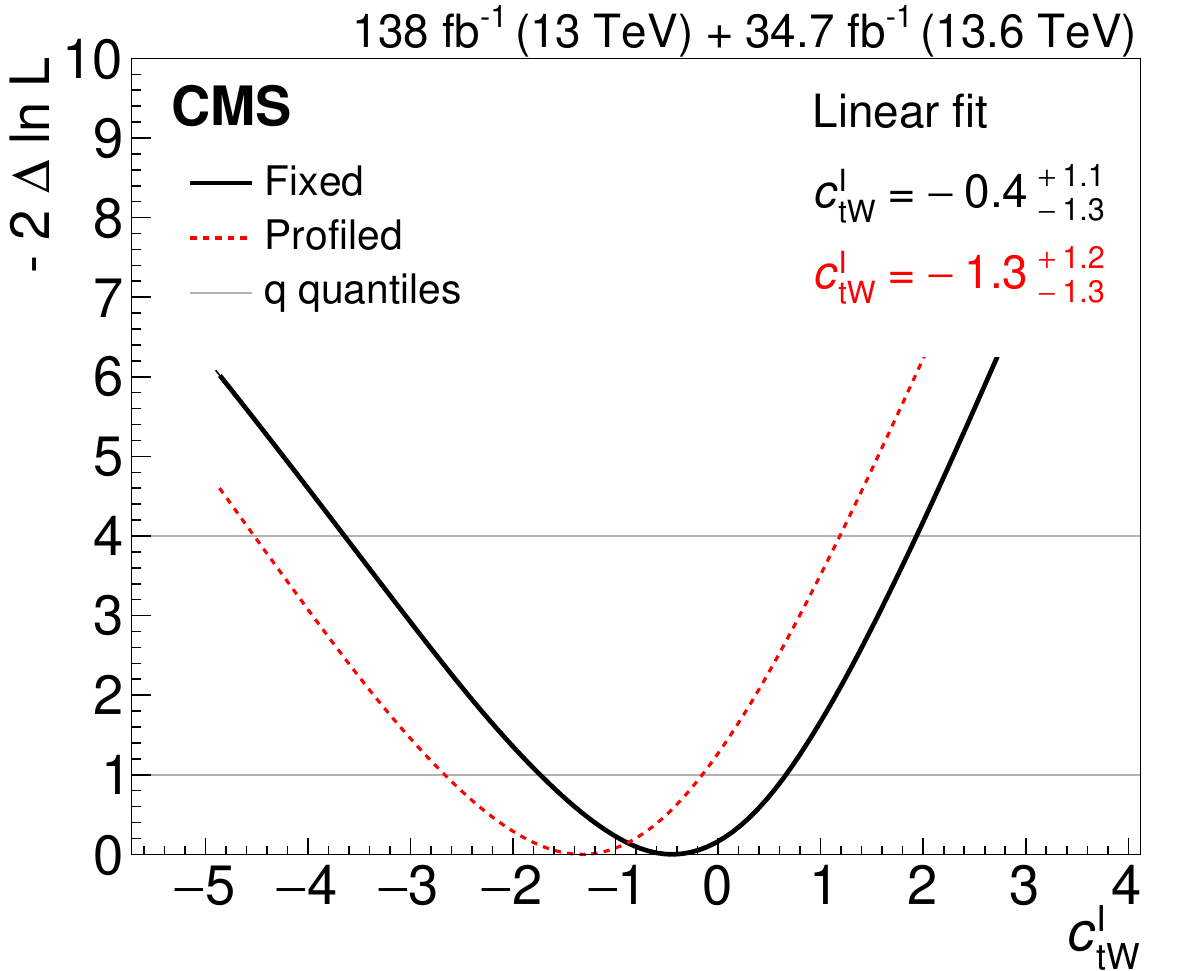}%
\hfill%
\includegraphics[width=0.47\textwidth]{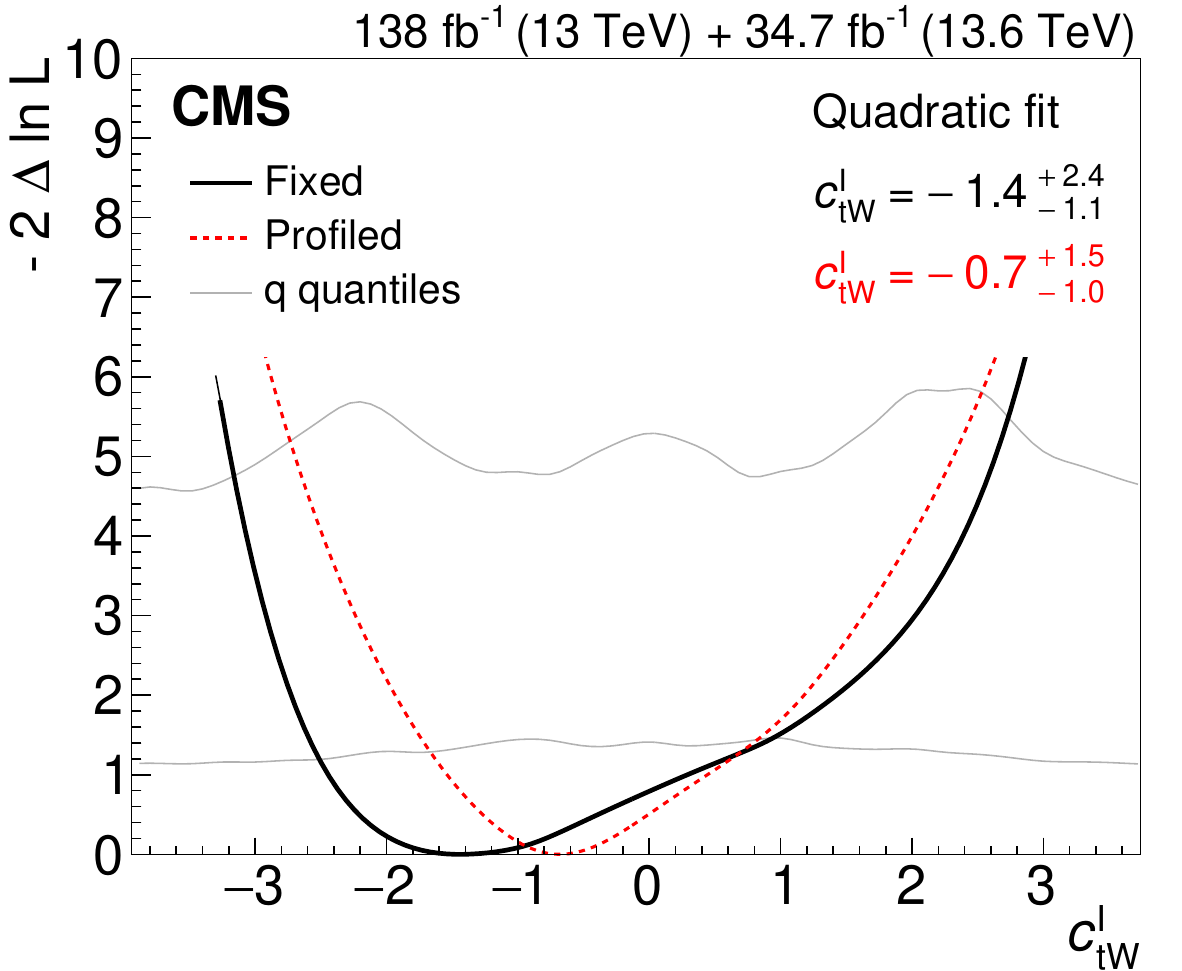} \\
\includegraphics[width=0.47\textwidth]{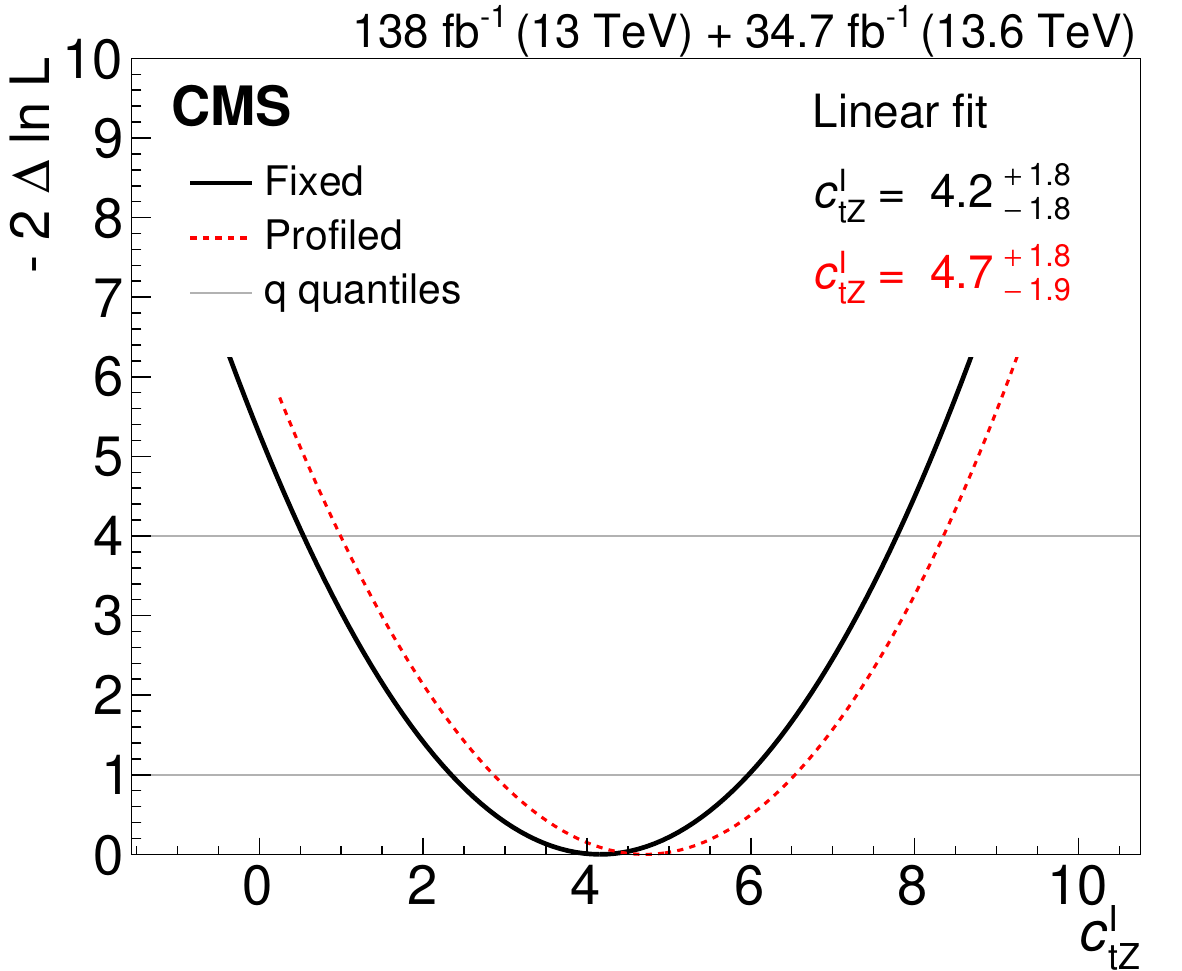}%
\hfill%
\includegraphics[width=0.47\textwidth]{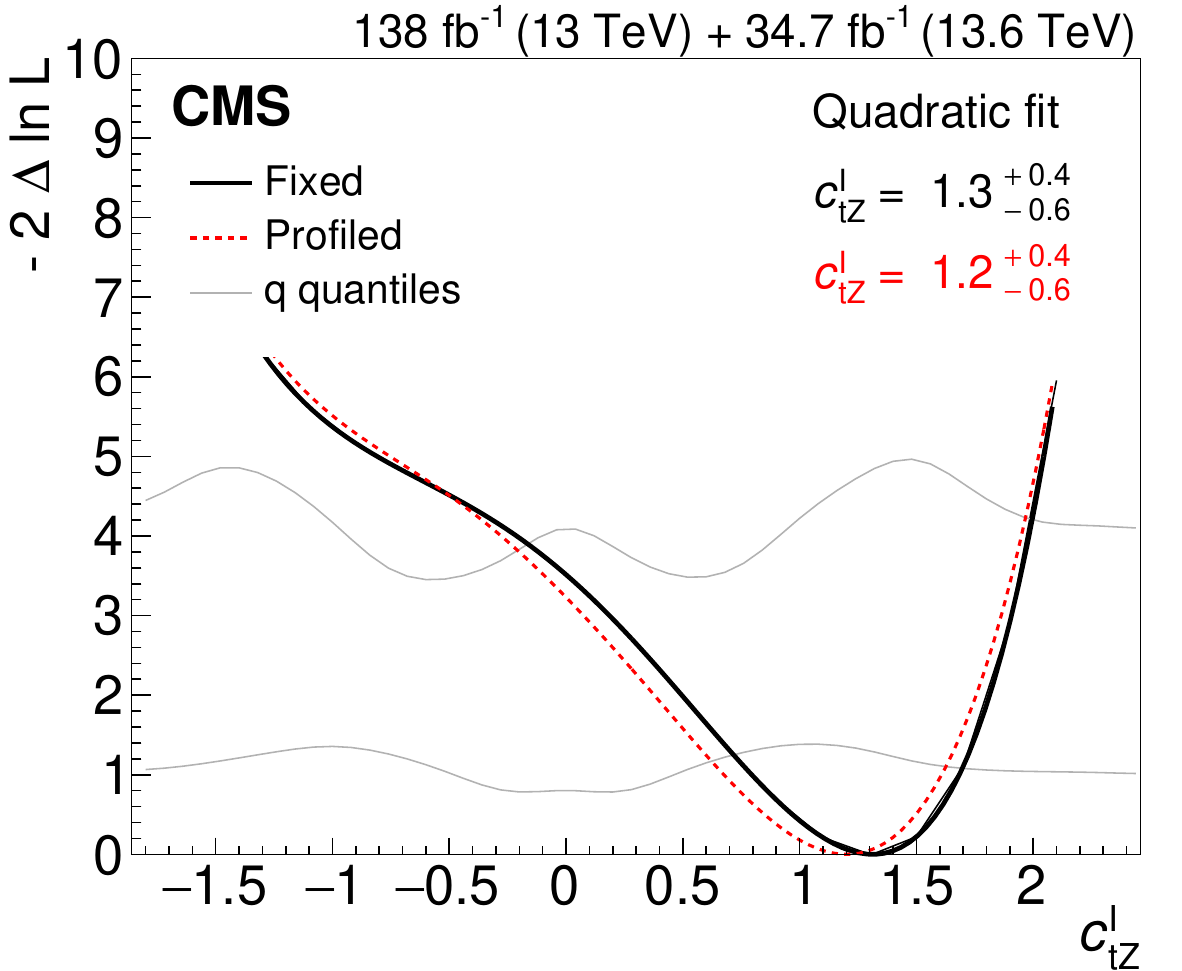}
\caption{%
    Likelihood scans as a function of \ctWI (upper row) and \ctZI (lower row), separately for cases in which the other coefficient is set to zero (dark black solid line) or profiled (red dashed line). Plots in the left (right) column represent the linear (quadratic) fit. Light gray lines represent the quantiles of the test statistics.
}
\label{fig:scans}
\end{figure*}

\begin{figure}[!htp]
\centering
\includegraphics[width=0.47\textwidth]{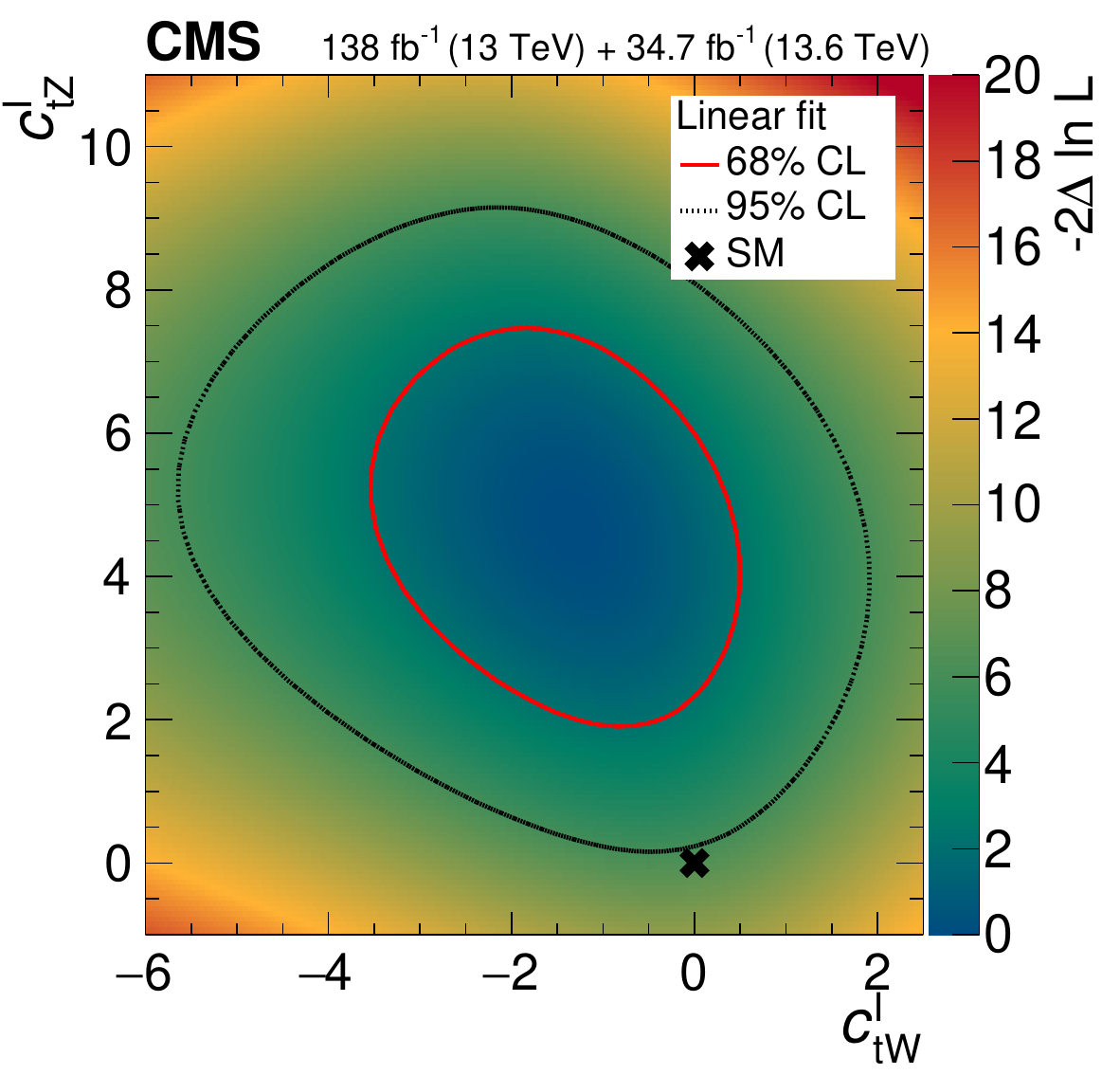}%
\hfill%
\includegraphics[width=0.47\textwidth]{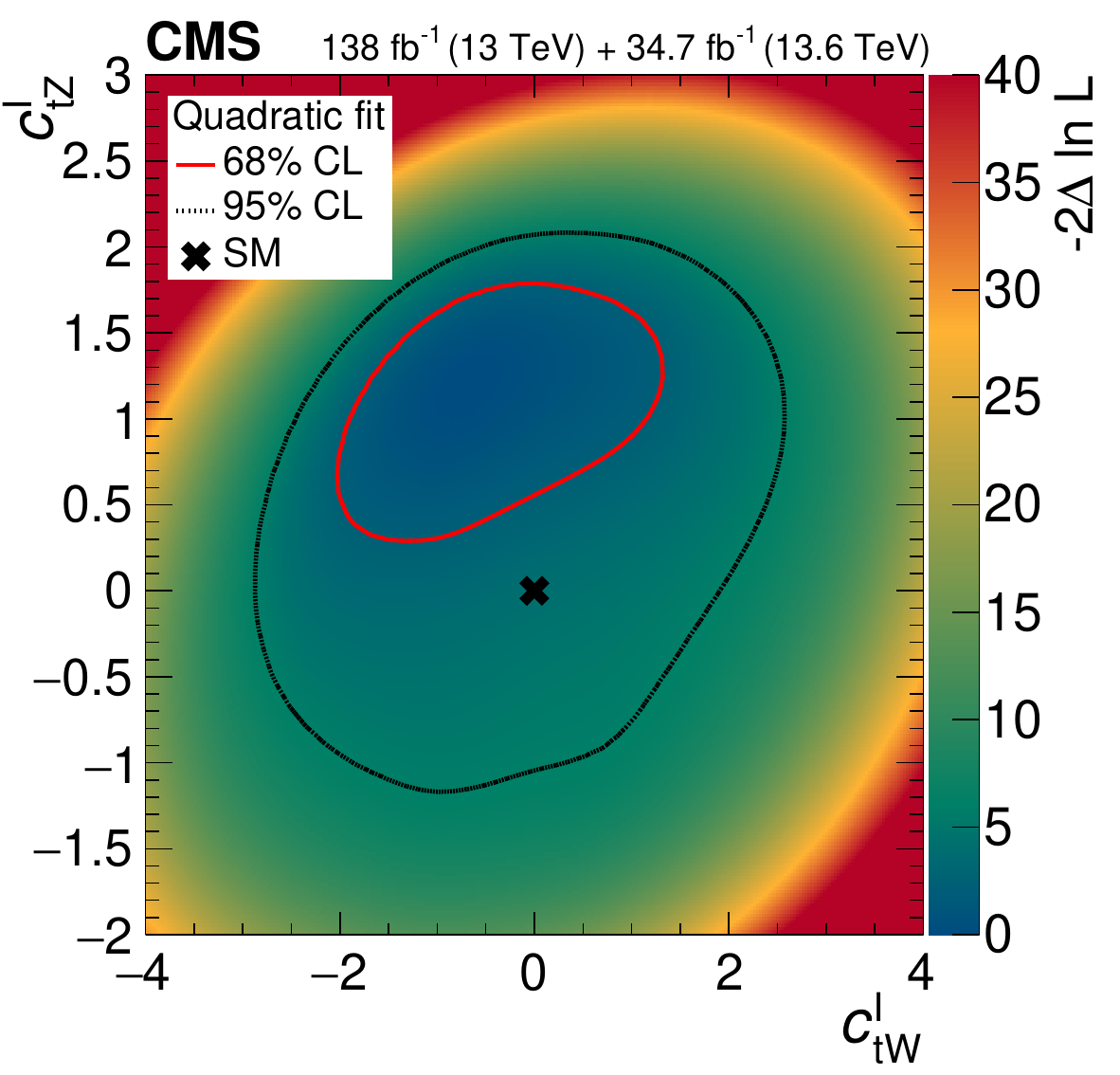}
\caption{%
    Likelihood scans as functions of \ctWI and \ctZI, including linear contributions only (\cmsleft)
    and both linear and quadratic contributions (\cmsright).
}
\label{fig:scans_2d}
\end{figure}

In order to perform inference on \ctWI and \ctZI, we compute the profile likelihood ratio test statistic
\begin{equation}
    q\Big(\ctWI,\ctZI\Big) = -2\ln \frac{\mathcal{L}\Big(\ctWI, \ctZI , \widehat{\widehat{\vectheta}}(\ctWI\,\ctZI) \Big)}{\mathcal{L}\Big(\widehat{\ctWI}, \widehat{\ctZI} , \widehat{\vectheta} \Big)}
\end{equation}
using the \textsc{combine} package~\cite{CMS:2024onh}
for different values of \ctWI and \ctZI, as shown in Figs.~\ref{fig:scans}--\ref{fig:scans_2d}. Here, $\widehat{\ctWI}$, $\widehat{\ctZI}$ and $\widehat{\vectheta}$ represent the maximum likelihood
estimate of the parameters, and $\widehat{\widehat{\vectheta}}(\ctWI,\ctZI)$ represents the maximum
likelihood estimation of \vectheta for given values of \ctWI and \ctZI.

We consider separately the case in which only the linear contribution in Eq.~\eqref{eq:eft_parameterization} is present (linear fit) and the case
in which the linear and quadratic terms contribute together (quadratic fit). In the linear fit, the analysis
is only sensitive to asymmetries in the observables, because the linear contributions do not
modify the total event yield or introduce effects that are symmetric on the observables. As a consequence,
such analysis is robust to the modeling of the signal and background processes, including their assumed cross sections. In the quadratic fits,
the sensitivity comes, in addition, from effects in the total number of events and symmetric shape variations
introduced by the quadratic contributions, as shown in Fig.~\ref{fig:distribution}. Quadratic fits are more sensitive to the
assumed \ttZ and \tZq total cross section, as well as the rest of the theoretical and experimental systematic uncertainties,
that have a negligible contribution to the result of the linear fit.
We also consider separately the case in which only one coefficient is allowed to take values different
from zero and the case where both coefficients are allowed to take nonzero values.

We use the scans in Figs.~\ref{fig:scans}--\ref{fig:scans_2d} to determine 68 and 95\% confidence level (\CL) intervals for individual and simultaneous coefficient measurements, respectively.
We employ the asymptotic approximation~\cite{Cowan:2010js} in the linear case. In the quadratic fits,
the asymptotic approximation does not hold~\cite{Bernlochner:2022oiw},
because the signal model cannot account for arbitrarily large downward fluctuations in the number of observed events with respect to the SM predictions.
As a consequence, there is an effective boundary in the statistical model on which the best-fit point may lie and
which contradicts the hypotheses required by the asymptotic approximation.
To set limits in a rigorous way,
we derive the distribution of $q\Big(\ctWI,\ctZI\Big)$ using pseudoexperiments for a range
of \ctWI and \ctZI values, whose quantiles are shown in Fig.~\ref{fig:scans}.
We note that the asymptotic approximation would lead to slightly
tighter constraints on \ctWI and \ctZI than our results.

The results show an overall good agreement between the measured \ctWI and \ctZI values and the SM prediction.
We use these results to set observed limits on \ctWI and \ctZI, and compare with the expected results obtained assuming an observation equal to the
SM hypothesis.
We set observed (expected) limits of  $\ctWIdown<\ctWI<\ctWIup$ and $\ctZIdown<\ctZI<\ctZIup$
($-2.0<\ctWI<2.0$ and $-1.5<\ctZI<1.5$) at 95\% \CL,
when considering the effect of both linear and quadratic terms, and the other coefficient is profiled. When considering only linear terms and profiling the other coefficient, we set observed (expected) limits of $\ctWIdownlinear<\ctWI<\ctWIuplinear$ and $\ctZIdownlinear<\ctZI<\ctZIuplinear$ ($-2.7<\ctWI<2.7$ and $-3.6< \ctZI<3.6$).
Simultaneous measurements of the two coefficients are compatible with the SM prediction within two standard deviations
in both the linear and quadratic cases.

When considering individual measurements of the two coefficients, the largest deviation is observed in \ctZI,
in which data is consistent with positive values of \ctZI. The significance of the discrepancy ranges
between 1.9 and 2.5 standard deviations, depending on whether quadratic contributions are considered and whether \ctWI is profiled
or fixed to its SM value. The deviation is associated with the asymmetry in the $g_{\ctZI}$ distribution and the
slight excess of observed events discussed earlier, both seen in Fig.~\ref{fig:stack}. As an additional check of the robustness of the result,
we performed a likelihood fit categorizing events in the different data-taking eras and allowing \ctZI to take
a different value in each era. We found that the \ctZI values obtained in the different eras were consistent within the uncertainties of the model.

The above results set limits for the first time on the interference between \ctZI and the SM contribution
and have competitive sensitivity to the interference between \ctWI and the SM contribution with respect to Ref.~\cite{ATLAS:2022vym},
which studies this operator in $t$-channel single top quark production. When considering also quadratic contributions, our measurement
reaches a comparable sensitivity to other searches studying these operators using the distribution of \CP-even observables in \ttZ and \ttA events~\cite{CMS:2019too, ATLAS:2024hmk},
and adds sensitivity to the sign of \ctZI that can only be probed with \CP-sensitive observables.

\section{Summary}

We have presented a search for additional sources of the charge-parity (\CP) symmetry violation in the
associated production of top quarks and a \PZ boson, in particular, for \ttZ and \tZq production
in final states with at least three leptons. The search is performed in a sample of proton-proton collision
data at center-of-mass energies of 13 and 13.6\TeV corresponding to a total integrated luminosity of  173\fbinv.
The measurement uses, for the first time in this topology, \CP-odd observables which are
constructed using physics-informed machine-learning techniques.
These observables are predicted by the standard model (SM) to be symmetrically distributed whereas
asymmetries could arise from \CP-violating effects.
The results are generally
consistent with the SM prediction. We set limits on the Wilson coefficients \ctWI and \ctZI
of $\ctWIdown<\ctWI<\ctWIup$ and $\ctZIdown<\ctZI<\ctZIup$, respectively, at 95\% confidence level, considering the effect of both linear and quadratic terms, and profiling the other coefficient. The largest discrepancy is observed in \ctZI, for which data is consistent with positive values.
When considering only linear terms, the observed local significance with respect to the SM hypothesis corresponds to 2.5 standard deviations.

\begin{acknowledgments}
We congratulate our colleagues in the CERN accelerator departments for the excellent performance of the LHC and thank the technical and administrative staffs at CERN and at other CMS institutes for their contributions to the success of the CMS effort. In addition, we gratefully acknowledge the computing centers and personnel of the Worldwide LHC Computing Grid and other centers for delivering so effectively the computing infrastructure essential to our analyses. Finally, we acknowledge the enduring support for the construction and operation of the LHC, the CMS detector, and the supporting computing infrastructure provided by the following funding agencies: SC (Armenia), BMBWF and FWF (Austria); FNRS and FWO (Belgium); CNPq, CAPES, FAPERJ, FAPERGS, and FAPESP (Brazil); MES and BNSF (Bulgaria); CERN; CAS, MoST, and NSFC (China); MINCIENCIAS (Colombia); MSES and CSF (Croatia); RIF (Cyprus); SENESCYT (Ecuador); ERC PRG, RVTT3 and MoER TK202 (Estonia); Academy of Finland, MEC, and HIP (Finland); CEA and CNRS/IN2P3 (France); SRNSF (Georgia); BMBF, DFG, and HGF (Germany); GSRI (Greece); NKFIH (Hungary); DAE and DST (India); IPM (Iran); SFI (Ireland); INFN (Italy); MSIP and NRF (Republic of Korea); MES (Latvia); LMTLT (Lithuania); MOE and UM (Malaysia); BUAP, CINVESTAV, CONACYT, LNS, SEP, and UASLP-FAI (Mexico); MOS (Montenegro); MBIE (New Zealand); PAEC (Pakistan); MES and NSC (Poland); FCT (Portugal); MESTD (Serbia); MICIU/AEI and PCTI (Spain); MOSTR (Sri Lanka); Swiss Funding Agencies (Switzerland); MST (Taipei); MHESI and NSTDA (Thailand); TUBITAK and TENMAK (Turkey); NASU (Ukraine); STFC (United Kingdom); DOE and NSF (USA).

\hyphenation{Rachada-pisek} Individuals have received support from the Marie-Curie program and the European Research Council and Horizon 2020 Grant, contract Nos.\ 675440, 724704, 752730, 758316, 765710, 824093, 101115353, 101002207, 101001205, and COST Action CA16108 (European Union); the Leventis Foundation; the Alfred P.\ Sloan Foundation; the Alexander von Humboldt Foundation; the Science Committee, project no. 22rl-037 (Armenia); the Fonds pour la Formation \`a la Recherche dans l'Industrie et dans l'Agriculture (FRIA-Belgium); the Beijing Municipal Science \& Technology Commission, No. Z191100007219010, the Fundamental Research Funds for the Central Universities, the Ministry of Science and Technology of China under Grant No. 2023YFA1605804, and the Natural Science Foundation of China under Grant No. 12061141002 (China); the Ministry of Education, Youth and Sports (MEYS) of the Czech Republic; the Shota Rustaveli National Science Foundation, grant FR-22-985 (Georgia); the Deutsche Forschungsgemeinschaft (DFG), among others, under Germany's Excellence Strategy -- EXC 2121 ``Quantum Universe" -- 390833306, and under project number 400140256 - GRK2497; the Hellenic Foundation for Research and Innovation (HFRI), Project Number 2288 (Greece); the Hungarian Academy of Sciences, the New National Excellence Program - \'UNKP, the NKFIH research grants K 131991, K 133046, K 138136, K 143460, K 143477, K 146913, K 146914, K 147048, 2020-2.2.1-ED-2021-00181, TKP2021-NKTA-64, and 2021-4.1.2-NEMZ\_KI-2024-00036 (Hungary); the Council of Science and Industrial Research, India; ICSC -- National Research Center for High Performance Computing, Big Data and Quantum Computing and FAIR -- Future Artificial Intelligence Research, funded by the NextGenerationEU program (Italy); the Latvian Council of Science; the Ministry of Education and Science, project no. 2022/WK/14, and the National Science Center, contracts Opus 2021/41/B/ST2/01369 and 2021/43/B/ST2/01552 (Poland); the Funda\c{c}\~ao para a Ci\^encia e a Tecnologia, grant CEECIND/01334/2018 (Portugal); the National Priorities Research Program by Qatar National Research Fund; MICIU/AEI/10.13039/501100011033, ERDF/EU, "European Union NextGenerationEU/PRTR", and Programa Severo Ochoa del Principado de Asturias (Spain); the Chulalongkorn Academic into Its 2nd Century Project Advancement Project, and the National Science, Research and Innovation Fund via the Program Management Unit for Human Resources \& Institutional Development, Research and Innovation, grant B39G670016 (Thailand); the Kavli Foundation; the Nvidia Corporation; the SuperMicro Corporation; the Welch Foundation, contract C-1845; and the Weston Havens Foundation (USA).
\end{acknowledgments}

\bibliography{auto_generated}
\cleardoublepage \appendix\section{The CMS Collaboration \label{app:collab}}\begin{sloppypar}\hyphenpenalty=5000\widowpenalty=500\clubpenalty=5000\cmsinstitute{Yerevan Physics Institute, Yerevan, Armenia}
{\tolerance=6000
A.~Hayrapetyan, V.~Makarenko\cmsorcid{0000-0002-8406-8605}, A.~Tumasyan\cmsAuthorMark{1}\cmsorcid{0009-0000-0684-6742}
\par}
\cmsinstitute{Institut f\"{u}r Hochenergiephysik, Vienna, Austria}
{\tolerance=6000
W.~Adam\cmsorcid{0000-0001-9099-4341}, J.W.~Andrejkovic, L.~Benato\cmsorcid{0000-0001-5135-7489}, T.~Bergauer\cmsorcid{0000-0002-5786-0293}, M.~Dragicevic\cmsorcid{0000-0003-1967-6783}, C.~Giordano, P.S.~Hussain\cmsorcid{0000-0002-4825-5278}, M.~Jeitler\cmsAuthorMark{2}\cmsorcid{0000-0002-5141-9560}, N.~Krammer\cmsorcid{0000-0002-0548-0985}, A.~Li\cmsorcid{0000-0002-4547-116X}, D.~Liko\cmsorcid{0000-0002-3380-473X}, M.~Matthewman, I.~Mikulec\cmsorcid{0000-0003-0385-2746}, J.~Schieck\cmsAuthorMark{2}\cmsorcid{0000-0002-1058-8093}, R.~Sch\"{o}fbeck\cmsAuthorMark{2}\cmsorcid{0000-0002-2332-8784}, D.~Schwarz\cmsorcid{0000-0002-3821-7331}, M.~Shooshtari, M.~Sonawane\cmsorcid{0000-0003-0510-7010}, W.~Waltenberger\cmsorcid{0000-0002-6215-7228}, C.-E.~Wulz\cmsAuthorMark{2}\cmsorcid{0000-0001-9226-5812}
% ...
\par}
\vskip\cmsinstskip
\dag:~Deceased\\
$^{1}$Also at Yerevan State University, Yerevan, Armenia\\
$^{2}$Also at TU Wien, Vienna, Austria\\
% ...
\end{sloppypar}
\end{document}